# ULXs, Microblazars, and the Unidentified EGRET sources

Yousaf M. Butt[1], Gustavo E. Romero[2], Diego F. Torres[3]

[1]*Harvard-Smithsonian Center for Astrophysics, 60 Garden St., Cambridge, MA, USA*

[2]*Instituto Argentino de Radioastronomía, CC 5, 1894, Villa Elisa, Buenos Aires, ARGENTINA*

[3] *Lawrence Livermore National Laboratory, 7000 East Ave., L-413, Livermore, CA 94551, USA*

**We suggest that ultraluminous X-ray sources (ULXs) and some of the variable low latitude EGRET gamma-ray sources may be two different manifestations of the same underlying phenomena: high-mass microquasars with relativistic jets forming a small angle with the line of sight (i.e. microblazars). Microblazars with jets formed by relatively cool plasma (Lorentz factors for the leptons up to a few hundreds) naturally lead to ULXs. If the jet contains very energetic particles (high-energy cutoff above Lorentz factors of several thousands) the result is a relatively strong gamma-ray source. As pointed out by Kaufman Bernadó, Romero & Mirabel (2002), a gamma-ray microblazar will always have an X-ray counterpart (although it might be relatively weak), whereas X-ray microblazars might have no gamma-ray counterparts.**

X-ray observations of external galaxies have revealed a population of point-like X-ray sources which are typically distributed near star-forming regions and which display considerable variability over ~month timescales. At their brightest such sources correspond to an *isotropic* luminosity of $\sim 10^{41}$ ergs/sec, and have been termed Ultra Luminous X-ray sources or ULXs (eg. Makishima et al., 2000). With the CHANDRA X-ray observatory it has been possible to simultaneously confirm their point-like character down to sub-arcsecond scales, and, in some cases, to measure their spectra. Various theories have been advanced to explain these sources, though it now appears that accretion onto a compact object must play a role in order to explain their high X-ray variability.

Kaaret et al. (2003) have recently observed radio emission from one such ULX in NGC 5408, which is consistent with relativistic jet emission expected from an accreting stellar mass black hole whose jet axis is beamed towards us. [Of course, if such beaming operates then the intrinsic luminosity of the ULXs can be much lower than the isotropic luminosity – by factors of $\sim 10^6$ (eg. Georganopoulos, Kirk & Mastichiadis, 2001).] Indeed, such stellar-mass 'microblazars' had already been predicted to exist by Mirabel & Rodriguez (1999). That the spectral state changes observed in at least some ULXs are reminiscent of those seen in stellar mass black-hole X-ray binaries (BHXB) supports this picture (eg. La Parola et al., 2001). Thus, a direct connection of the extragalactic ULXs with the Galactic microblazar candidate sources (eg. V4641Sgr; Orosz et al., 2001) seems quite plausible and even likely (King et al., 2001; Körding, Falcke & Markoff, 2002; Kaaret et al. 2003).

The mechanism thought to produce the beamed X-ray emission is Compton upscattering of photons from the companion star or accretion disk by the power-law distribution of $e^{\pm}$ plasma contained in the collimated, relativistic jet (eg. Georganopolous, Ahronian & Kirk 2002 -GAK; Band & Grindlay, 1986; but see also Aharonian & Atoyan, 1999 and Markoff, Falcke & Fender, 2001 for synchrotron models). GAK find that an upper-cutoff of the $e^{\pm}$ plasma power-law distribution at Lorentz factors of, $\gamma_2 \sim 100\text{-}200$, produces a strong non-thermal X-ray luminosity in high-mass microquasars through external Compton upscattering of the stellar UV field, even though their approach ignores the effects of the hot corona or ADAF region that surrounds the central compact object. The existence of this region has been established from the Compton reflection features observed in Cygnus X-1 and other objects (see, eg., Zdziarski et al., 2002). However, Romero, Kaufman Bernadó & Mirabel (2002) have found that even when this coronal emission and its interaction with the relativistic jet are taken into account, GAK's basic conclusion that IC emission from a microblazar jet with $\gamma_2 \sim 100\text{-}200$ suitably reproduces a ULX spectrum, remains valid.

If the upper-cutoff of the $e^{\pm}$ plasma Lorentz factor, $\gamma_2 \gtrsim 10^3$, then instead of a beamed X-ray source what will be observed is a gamma-ray source (Kaufman Bernadó, Romero & Mirabel, 2002 -KRM), though a weaker X-ray counterpart is also predicted. Whereas the gamma-ray emission is amplified by a factor $D^{2+p}=D^{3+2\alpha}$, the synchrotron jet emission is amplified only by $D^{2+\alpha}$ (Dermer, 1995), where $\alpha=(p-1)/2$ is the synchrotron spectral index and D is the normal Doppler factor $D=[\Gamma(1-\beta\cos\phi)]^{-1}$, given in terms of the jet bulk velocity ($\beta$), bulk Lorentz factor ($\Gamma$) and the viewing angle ($\phi$). Thus, 'red' microblazars, ie. whose synchrotron SED peaks at IR energies, will not be ULXs since only a minor component of the radiated power goes into the synchrotron tail of their spectra and, furthermore, this emission is less strongly boosted than the inverse Compton gamma-ray emission. Conversely, at low values of the maximum $e^{\pm}$ plasma Lorentz factor, $\gamma_2 \sim 10^2$ we expect an X-ray source of ($L_x \sim 10^{33-34}$ ergs sec$^{-1}$) without a gamma-ray counterpart (KRM).

Thus, Galactic stellar-mass gamma-ray blazars may well exist – indeed, they may *already* have been detected by the EGRET instrument as a handful of the low-latitude variable MeV-GeV unidentified gamma-ray sources (Paredes et al, 2000; Romero, 2001; Grenier, 2001; Butt et al., 2002). We propose that there may exist a family of variable high-energy accreting beamed sources of stellar mass: those that are bright in X-rays (warm jets; $100 \lesssim \gamma_2 \lesssim 1000$), others in MeV, GeV and even higher energy gamma-rays (hot jets; $\gamma_2 \gtrsim 1000$).

The prime candidates for exploring the 'hot jet' microblazar hypothesis are the most variable EGRET (MeV-GeV range) gamma-ray sources (Hartman et al., 1999) observed at low Galactic latitudes, which show no positional coincidence with any known counterparts at other frequencies (Torres et al., 2001a). We list the best candidates below:

| Variable $\gamma$-Source | l | b | $\Delta\theta$ | $\Gamma$ | $F\gamma \times 10^{-8}$ | $I^*$ | |
|---|---|---|---|---|---|---|---|
| 0416+3640[†] | 162.2 | -9.97 | 0.63 | 2.59±0.32 | 21.1 | 2.6 | *Information from Hartman et al., 1999* |
| 1704-4732 | 340.1 | -3.79 | 0.66 | 1.86±0.33 | 29.5 | 3.0 | |
| 1735-1500 | 10.73 | 9.22 | 0.77 | 3.24±0.47 | 19.0 | 8.9 | |
| 1744-3934 | 350.8 | -5.4 | 0.66 | 2.42±0.17 | 26.5 | 3.0 | |
| 1746-1001 | 16.34 | 9.64 | 0.76 | 2.55±0.18 | 29.7 | 3.2 | |
| 1810-1032 | 18.81 | 4.23 | 0.39 | 2.29±0.16 | 31.5 | 2.6 | |
| 1812-1316 | 16.7 | 2.39 | 0.39 | 2.29±0.11 | 43.0 | 2.6 | |
| 1828+0142 | 31.90 | 5.78 | 0.55 | 2.76±0.39 | 30.8 | 5.3 | |
| 1834-2803 | 5.92 | -8.97 | 0.52 | 2.62±0.20 | 17.9 | 2.8 | |
| 1904-1124 | 24.22 | -8.12 | 0.50 | 2.60±0.21 | 22.5 | 2.9 | |
| 1928+1733 | 52.91 | 0.07 | 0.75 | 2.23±0.32 | 38.6 | 4.0 | |
| 2035+4441 | 83.17 | 2.50 | 0.54 | 2.08±0.26 | 39.1 | 3.4 | |

*l,b* are the Galactic longitude and latitude; $\Delta\theta$ is the size of the gamma ray error box in degrees; $\Gamma$ is the high-energy spectral index; $F\gamma \times 10^{-8}$ is the mean weighted flux >100 MeV in units of $10^{-8}$ ph cm$^{-2}$ sec$^{-1}$; and I is the 'variability index' (see Torres et al., 2001a for more details).
*An alternative variability analysis by Tompkins (1999) also classify most of these sources as variable. For a comparative analysis of variability indicies see Torres et al., 2001b
[†]possible AGN (Hartman et al., 1999).

Interestingly, the MeV range variable gamma-ray source recently reported near the Galactic plane by COMPTEL, (*l,b*~311.5°, -2.5° ; Zhang, Collmar & Schönfelder, 2002), may be the signature of an intermediate-range Lorentz factor microblazar ($\gamma_2$~1000), neatly connecting the keV and GeV domains.

The proposed variations in the upper-cutoff of plasma distribution Lorentz factors may naturally arise in different microquasars as a consequence of their different abilities to transform the jet bulk motion into internal energy of the plasma. Since such transformations are usually accomplished by shocks, we speculate that there exists a natural dispersion in the strength of these internal jet shocks across the various microquasars, just as there is a dispersion in the jet bulk velocities. The reason for such dispersions are probably related to the nature and properties of the compact object as well as variations in jet plasma composition (eg. the baryonic fraction).

The proposed unification between beamed, variable, gamma- and X-ray emission from BHXB systems is important to explore observationally. As previously mentioned, even gamma-ray microblazars would be expected to have weak, but significant, X-ray counterparts of the order of ~$10^{33}$-$10^{34}$ ergs sec$^{-1}$ (Kaufman Bernadó, Romero & Mirabel, 2002): a case study has been presented already by Butt et al. (2002) for the 3EG 1828+0142 source. Simultaneous gamma-ray, X-ray and radio searches of the EGRET error boxes listed above may be useful to better understand the physics of such jet sources – of all flavors, Galactic and extragalactic.

The high sensitivity and resolution of the CHANDRA X-ray observatory and the VLA & ATCA are ideally suited to search for such weak flaring point-like X-ray and radio counterparts of these variable gamma-ray sources. Information from JEM-X and IBIS on INTEGRAL will also be very useful for following up on, and identifying new intermediate jet ($\gamma_2$~1000) cases. Indeed, the wide fields of view permitted by these coded mask detectors are especially well suited to exploring the, typically 0.5°×0.5° sized, EGRET error boxes. In the near future the GeV-range observatories, AGILE and GLAST, could sample the proposed 'hot jet' sources directly. The smaller gamma-ray error boxes expected from these new observatories (especially GLAST) would be ideally suited to simultaneous multiwavelength observations.

The work of DFT was performed under the auspices of the U.S. Department of Energy (NNSA) by University of California Lawrence Livermore National Laboratory under contract No. W-7405-Eng-48. DFT is Lawrence Fellow in Astrophysics. YMB acknowledges the support of the *CHANDRA* project, NASA Contract NAS8-39073. G.E.R. is supported by the research grants PICT 03-04881 (ANPCT) and PIP 0438/98 (CONICET), as well as by Fundación Antorchas.